\documentclass[aip,cha,english,reprint,a4paper]{revtex4-2}

\usepackage[T1]{fontenc}
\usepackage{graphicx}
\usepackage{dcolumn}
\usepackage{bm}

\usepackage{amsmath,amssymb}
\usepackage{mathptmx}
\usepackage{physics}
\usepackage{etoolbox}
\usepackage[caption = false]{subfig}
\usepackage{standalone}
\usepackage{xcolor}
\usepackage{enumitem}
\usepackage{mathtools}
\usepackage{makecell}
\usepackage[colorlinks = true,linkcolor = red,citecolor = magenta]{hyperref}
\usepackage[sort&compress]{natbib}

\newcommand{\pcsadd}{Center for Theoretical Physics of Complex Systems, Institute for Basic Science (IBS), Daejeon 34126, Republic of Korea}
\newcommand{\ustadd}{Basic Science Program, Korea University of Science and Technology (UST), Daejeon 34113, Republic of Korea}

\makeatletter
\def\@email#1#2{
  \endgroup
    \patchcmd{\titleblock@produce}
      {\frontmatter@RRAPformat}
      {\frontmatter@RRAPformat{\produce@RRAP{*#1\href{mailto:#2}{#2}}}\frontmatter@RRAPformat}
      {}{}
}%
\makeatother

\newcommand{\mh}{\mathcal{H}}
\newcommand{\mhfd}{\mh_\mathrm{FD}}
\newcommand{\mhsd}{\mh_\mathrm{SD}}
\newcommand{\mhabf}{\mh_\mathrm{ABF}}

\newcommand{\mbet}{\langle\tau\rangle}
\newcommand{\mbeet}{\langle\tau\rangle_\mathrm{e}}

\begin{document}

\preprint{AIP/123-QED}

\title[Critical States Generators from Perturbed Flatbands]{Critical States Generators from Perturbed Flatbands}

\author{S. Lee}
  \email{scott430@naver.com}
  \affiliation{\pcsadd}
  \affiliation{\ustadd}
    
\author{S. Flach}
  \email{sflach@ibs.re.kr}
  \affiliation{\pcsadd}
  \affiliation{\ustadd}

\author{Alexei Andreanov}
  \email{aalexei@ibs.re.kr}
  \affiliation{\pcsadd}
  \affiliation{\ustadd}

\date{\today}

\begin{abstract}
  One-dimensional all-bands-flat lattices are networks with all bands being flat and highly degenerate. 
  They can always be diagonalized by a finite sequence of local unitary transformations parameterized by a set of angles \(\theta_{i}\).
  In our previous work, Ref.~\onlinecite{lee2023critical}, we demonstrated that quasiperiodic perturbations of the one-dimensional all-bands-flat lattice with \(\theta_{i} = \pi/4\) give rise to a critical-to-insulator transition and fractality edges separating critical from localized states.
  In this study we consider the full range of angles \(\theta_{i}\)s available for the all-bands-flat model and study the effect of the quasiperiodic perturbation.
  For weak perturbation, we derive an effective Hamiltonian and we identify the sets of \(\theta_{i}\)s for which the effective model maps to extended or off-diagonal Harper models and hosts critical states.
  For all the other values of the angles the spectrum is localized.
  Upon increasing the perturbation strength, the extended Harper model evolves into the system with energy dependent critical-to-insulator transitions, that we dub \emph{fractality edges}.
  The case where the effective model maps onto the off-diagonal Harper model features a critical-to-insulator transition at a finite disorder strength.
\end{abstract}

\maketitle

\begin{quotation}
  The breaking of the macroscopic degeneracy in flatband systems allows to realize a variety of interesting and exotic phases depending on the types of the perturbation applied.
  One particular example is all-band-flat systems where all energy bands are flat.
  It requires high degree of fine-tuning, but in 1D is always achieved by applying local unitary transformations to decoupled sites producing the entire manifold of all-bands-flat Hamiltonians.
  In the previous work,~\cite{lee2023critical} we studied the effect of a quasiperiodic perturbation on a specific all-bands-flat Hamiltonian (a specific point of the manifold).
  We found a critical-to-insulator transition analytically for weak perturbation and fractality edges numerically for finite perturbation.
  \par
  In this work, we consider the full manifold of all-bands-flat systems in presence of the quasiperiodic perturbation.
  Considering the full ABF manifold might be relevant for possible experimental realizations of these perturbed models.
  We identify analytically submanifolds with critical eigenstates for weak potential strengths.
  For finite strengths we confirm numerically the emergence of fractality edges on these submanifolds.
  In all other cases the perturbed Hamiltonians have all their eigenstates localized.
\end{quotation}

\section{Introduction}

Recently, physical systems with macroscopic degeneracies have received a lot of attention.
Their study is motivated by the observation that macroscopic degeneracies are fragile and are easily lifted even by weak perturbations, resulting in various exotic and unusual correlated phases.
One example of such systems is a flatband Hamiltonian -- a translationally invariant tight-binding network with dispersionless Bloch bands \(E(\mathbf{k}) = E\).~\cite{maksymenko2012flatband,leykam2018artificial,rhim2021singular}
The fine-tuned geometry or the symmetry of the flatband system causes destructive interference which leads to zero group velocity \(\nabla_{\mathbf{k}}E\) and traps particles over a strictly finite number of sites,~\cite{sutherland1986localization,aoki1996hofstadter} 
resulting in the appearance of compact localized states -- eigenstates with strictly compact support.
The extreme sensitivity of macroscopically degenerate flatbands to perturbations leads to emergence of a variety of interesting and exotic phases:
flatband ferromagnetism,~\cite{tasaki1992ferromagnetism} frustrated magnetism,~\cite{ramirez1994strongly, derzhko2015strongly} unconventional Anderson localization,~\cite{flach2014detangling} ergodicity breaking~\cite{kuno2020flat,danieli2020many,vakulchyk2021heat,danieli2022many} and superconductivity.~\cite{cao2018unconventional}

Flatband systems can be further fine-tuned to turn all their energy bands flat, resulting in the \emph{all-bands-flat} (ABF) networks.~\cite{vidal1998aharonov,danieli2021nonlinear}
Despite the high degree of fine-tuning they can be constructed systematically in 1D via a sequence of local unitary transformations applied to isolated sites.~\cite{flach2014detangling,danieli2021nonlinear,danieli2020many,cadez2021metal,lee2023critical}.
We conjectured that this results also holds in higher dimensions.~\cite{danieli2021nonlinear}
ABF models are even more sensitive to perturbations and interesting phenomena have been observed in presence of perturbations:
nonperturbative metal-insulator transitions~\cite{cadez2021metal,longhi2021inverse,li2022aharonov}, ergodicity-breaking~\cite{kuno2020flat,danieli2020many,vakulchyk2021heat} and caging of particles.~\cite{tovmasyan2013geometry,tovmasyan2018preformed,danieli2021nonlinear,danieli2021quantum,swaminathan2023signatures}

In Ref.~\onlinecite{lee2023critical}, we looked at a specific two-band ABF model and studied its behavior in the presence of a \emph{quasiperiodic perturbation}.
We identified parameters of the model for which it features critical states with subdiffusive/almost-diffusive transport for weak perturbation and an exotic phase transition -- critical-to-insulator transition (CIT) -- was observed.
For finite potential strength, we discovered fractality edges, an energy dependent CITs.
In this paper, we explore the full manifold of this two-band ABF ladder and identify additional submanifolds which also host critical states, CIT and fractality edges under the quasiperiodic perturbation.

The outline of the paper is the following:
We discuss the construction of the ABF in Sec.~\ref{sec:model}.
Then we derive an effective model in Sec.~\ref{sec:weak} valid for weak quasiperiodic perturbation and use it to locate the ABF submanifolds supporting critical states.
The properties of the full model at finite perturbation strengths are studied in Sec.~\ref{sec:finite}, followed by conclusions in Sec.~\ref{sec:conclusion}.

\section{The model}
\label{sec:model}

\begin{figure}
\centering
  \includegraphics[width = \columnwidth]{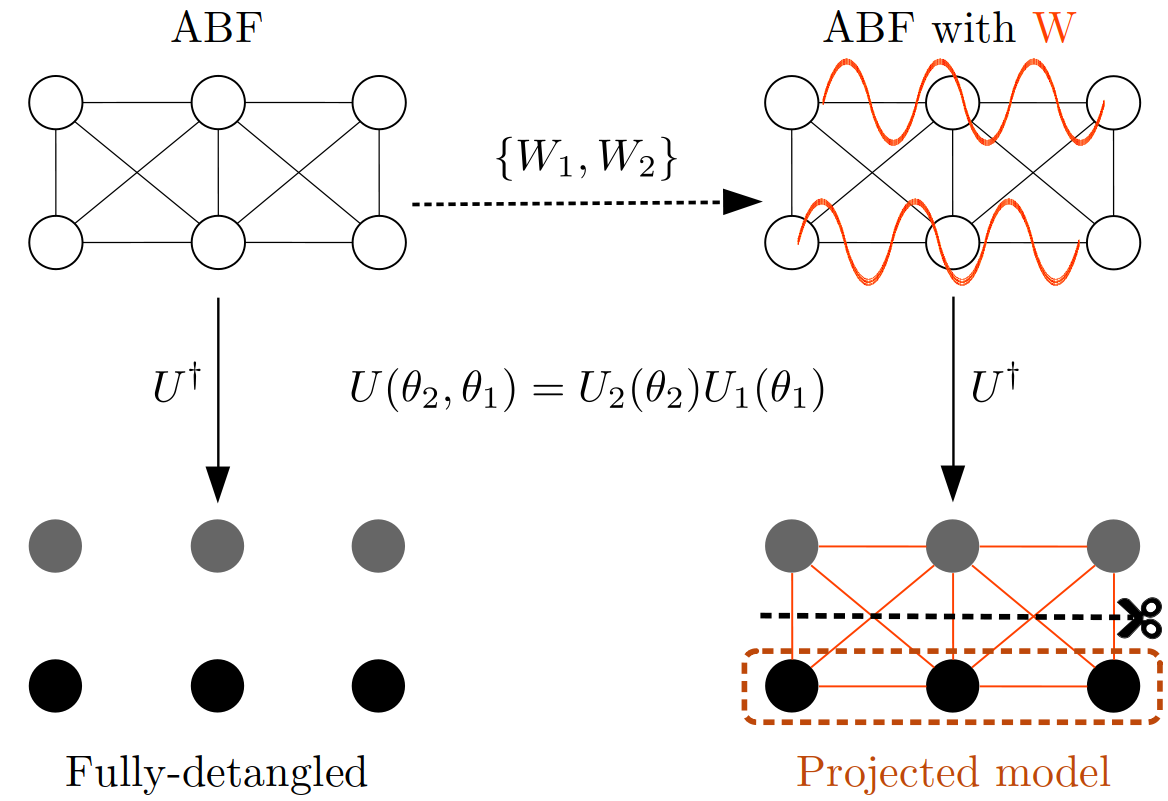}
  \caption{
    The ABF model is fully detangled/diagonalized into a diagonal Hamiltonian via local unitary transformations \(U_{1}\) and \(U_{2}\) with appropriate angles \(\theta_{1}\) and \(\theta_{2}\) (left).
    By adding quasiperiodic perturbation \(W\) made up of two quasiperiodic fields \(W_{1,2}\) in Eq.~\eqref{eq:fields}, nontrivial hoppings are created in the fully detangled basis (right).
    For weak quasiperiodic fields, the first-order degenerate perturbation theory is used to derive an effective projected model.
    In the detangled basis, this corresponds to only keeping the hopping terms coupling the sites on the same sublattices.
  }
  \label{fig:scheme}
\end{figure}

The model we focus on is the one-dimensional (1D) ABF ladder with two flatbands and the nearest-neighbor unit cells hopping,~\cite{maimaiti2019universal} summarized in Fig.~\ref{fig:scheme}.
Its Hamiltonian \(\mhabf\) is constructed from a macroscopically degenerate diagonal matrix \(\mhfd\) with onsite energies \(\varepsilon_a\) and \(\varepsilon_b\) on the two sublattices:~\cite{danieli2021nonlinear}
\begin{gather}
  \mhfd = \sum \varepsilon_{a}\ketbra{a_{n}}{a_{n}} + \varepsilon_{b}\ketbra{b_{n}}{b_{n}},
\end{gather}
where \(a_n, b_n\) are the basis states of the \(2\) sublattices and the bandgap \(\Delta = \abs{\varepsilon_a - \varepsilon_b}\).
We term \(\mhfd\) as the parent Hamiltonian for the manifold of ABF systems and refer to it as \emph{fully detangled}.~\cite{flach2014detangling,danieli2021nonlinear,danieli2020many}

The construction of the ABF manifold is based on a sequence of unitary transformations \(U_i\)~\cite{flach2014detangling,danieli2021nonlinear,danieli2020many,cadez2021metal,lee2023critical} applied to the parent Hamiltonian.~\footnote{
We note the relative similarity of this procedure with the construction of the Haar random unitary matrices.~\cite{zyczkowski1994random}
}
Each transformation \(U_i\) is in turn a direct sum of local unitary transformations, which need not commute for different \(i\).
Throughout the paper, ABF Hamiltonian always refers to a connected ABF network.
For a one-dimensional system with nearest-neighbor unit cell hopping only two local unitary transformations \(U=U_2 U_1\) are enough to produce a connected hopping network.
Each transformation \(U_{1,2}\) is a direct sum of local transformations, e.g. it takes a block diagonal form.
The most general \(2\times 2\) block is a \(\mathrm{SU}(2)\) matrix:
\begin{align}
  U_{i} = \sum_{n, n^{\prime}\in\mathbb{N}}
  z_{i}\vert{a^{(i)}_{n}}\rangle\langle{a^{(i-1)}_{n}}\vert &+ w_{i}\vert{a^{(i)}_{n}}\rangle\langle{b^{(i-1)}_{{n}^{\prime}}}\vert \notag \\
  - w_{i}^{*}\vert{b^{(i)}_{n}}\rangle\langle{a^{(i-1)}_{{n}^{\prime}}}\vert &+ z_{i}^{*}\vert{b^{(i)}_{n}}\rangle\langle{b^{(i-1)}_{n}}\vert. 
  \label{eq:unitary}
\end{align}
The index \(i\) denotes the \(i\)-th local unitary transformation and the indices \(n, n^\prime\) label the unit cells.
In the simplest case, the blocks are parameterized by only two angles: \(\theta_{1,2}\) for \(U_{1,2}\), respectively, producing real \(U_{1,2}\in\mathrm{SO(2)}\).
The \(2\times 2\) blocks can act on the \(2\) sites within the same unit cell, \(n^\prime = n\), or different unit cells, \(n^\prime \neq n\).
This generates different, non-commuting transformations.~\cite{danieli2021nonlinear,danieli2020many,cadez2021metal}
We choose the \(U_1 \in \mathrm{SO(2)}\) blocks to act within the same unit cell (\(n^\prime=n\) in the above expression), 
while for the \(U_2 \in \mathrm{SO(2)}\) blocks, one of the sublattice sites is taken from the neighboring unit cell: \(n^\prime = n - 1\).
As was discussed in Ref.~\onlinecite{lee2023critical}, one can consider the angles \(0\leq\theta_{1,2}\leq\pi/2\) only, thanks to the symmetries of the model. 

A quasiperiodic perturbation \(W\) is added to the ABF Hamiltonian, \(\mh = \mhabf + W\).
It is a diagonal matrix depending ony two quasiperiodic fields \(W_{1}\) and \(W_{2}\) defined as follows:
\begin{align}
    \label{eq:fields}
    W_{1}(n) &= \lambda_{1}\cos(2\pi\alpha n + \phi), \notag \\
    W_{2}(n) &= \lambda_{2}\cos(2\pi\alpha n + \beta + \phi). 
\end{align}
\(W_{1,2}\) are applied to the \(2\) sublattices of the ladder respectively, see Fig.~\ref{fig:scheme}.
The spatial frequency \(\alpha\) is an irrational number, \(\alpha \in \mathbb{R}\setminus\mathbb{Q}\), \(\beta\) is the phase difference between \(W_{2}\) and \(W_{1}\),
while \(\phi\) is the phase used for averaging of the results, and \(\lambda_{1,2}\) are the strengths of the quasiperiodic potentials.

Our results are summarized in Table~\ref{tab:weak-table}.
For parameters not listed in the table, all the eigenstates are always localized.

\begin{table}
    \centering
    \setlength{\tabcolsep}{5pt}
    \renewcommand{\arraystretch}{3}
    \begin{tabular}{c|c|c||c}
      \noalign{\smallskip}\noalign{\smallskip}\hline\hline
      Submanifold & \makecell{Weak \\ perturbation} & \makecell{Type of\\ eigenstates and\\ wavespreading} & \makecell{Finite\\ perturbation} \\
      \hline
      \makecell{Any \(\beta\) \\ \(\lambda_{2}/\lambda_{1} = 0\) \\ \(\theta_{1} + \theta_{2} = \pi/2\)} & EHM & \makecell{Critical \\ and \\ subdiffusive} & CIT \\
      \hline
      \makecell{\(\beta = \pi\) \\ \(\lambda_{2}/\lambda_{1} = 1\) \\ \(\forall\theta_{1}: \theta_{2} = \pi/4\)} & OHM & \makecell{Critical \\ and \\ almost diffusive} & \makecell{Perturbation \\ independent \\ fractality edges} \\
      \hline\hline
    \end{tabular}
    \caption{
      Summary of the eigenstates properties for the projected model in Eq.~\eqref{eq:pm} and for the finite perurbation strength.
      EHM and OHM stand for the extended and the off-diagonal Harper model, respectively.
      They have qualitatively different transport properties.
    }
    \label{tab:weak-table}
\end{table}

\section{Weak quasiperiodic potential and effective projected model}
\label{sec:weak}

We start by analyzing the limit of vanishing/weak \(\lambda_{1,2}\) compared to the bandgap \(\Delta = \abs{\varepsilon_a - \varepsilon_b}\).
In this limit, we can simplify the model by using the first order degenerate perturbation theory.
First, we apply the inverse unitary transformation \(U^{\dagger}\) on \(\mh\) to go back to the originally fully-detangled system, \(\tilde{\mh} = \mhfd + U^{\dagger} W U\).
However, due to the perturbation \(W\), there are now additional hoppings given by \(U^{\dagger} W U\) while \(\mhfd\) acts merely as a sublattice dependent onsite potential.
Second, we use perturbation theory~\cite{cadez2021metal,lee2023critical} to project the Hamiltonian onto one of the sublattices: this amounts to neglecting all the terms related to the other sublattice.
A similar procedure can be used for other perturbations, including interactions.~\cite{tovmasyan2013geometry,takayoshi2013phase,pelegri2020interaction,kuno2020interaction}
The procedure is outlined schematically in Fig.~\ref{fig:scheme}.
This produces an effective one-dimensional tight binding problem which is referred to as the \emph{projected model}:~\cite{cadez2021metal, lee2023critical}
\begin{gather}
    \label{eq:pm}
    E a_{n} = v_{n}a_{n} + t_{n-1}a_{n-1} + t_{n}a_{n+1}.
\end{gather}
The quasiperiodicity of the original potentials \(W_{1,2}\) is reflected in the quasiperiodicity of the onsite potential \(v_{n}\) and the hopping \(t_{n}\):
\begin{align}
    \label{eq:coefficients:v}
    v_{n} &= v_{s}\sin(2\pi\alpha n - \pi\alpha) + v_{c}\cos(2\pi\alpha n - \pi\alpha), \\
    \label{eq:coefficients:t}
    t_{n} &= t_{s}\sin(2\pi\alpha n) + t_{c}\cos(2\pi\alpha n).
\end{align}
The amplitudes \(v_{s,c}\) and \(t_{s,c}\) of \(v_{n}\) and \(t_{n}\) are given by:
\begin{align}
  v_{s} &= \sin(\pi\alpha)\left(\sin^2\theta_{1}\sin^2\theta_{2} - \cos^2\theta_{1}\cos^2\theta_{2}\right) \vphantom{\frac{\lambda_{2}}{\lambda_{1}}} \\
  &+ \frac{\lambda_{2}\cos\beta}{\lambda_{1}}\sin(\pi\alpha)\left(\sin^2\theta_{1}\cos^2\theta_{2} - \cos^2\theta_{1}\sin^2\theta_{2}\right) \notag \\
  &- \frac{\lambda_{2}\sin\beta}{\lambda_{1}}\cos(\pi\alpha)\left(\cos^2\theta_{1}\sin^2\theta_{2} + \sin^2\theta_{1}\cos^2\theta_{2}\right) \notag \\
  v_{c} &= \cos(\pi\alpha)\left(\cos^2\theta_{1}\cos^2\theta_{2} + \sin^2\theta_{1}\sin^2\theta_{2}\right) \vphantom{\frac{\lambda_{2}}{\lambda_{1}}} \\
  &+ \frac{\lambda_{2}\sin\beta}{\lambda_{1}}\sin(\pi\alpha)\left(\sin^2\theta_{1}\cos^2\theta_{2} - \cos^2\theta_{1}\sin^2\theta_{2}\right) \notag \\
  &+ \frac{\lambda_{2}\cos\beta}{\lambda_{1}}\cos(\pi\alpha)\left(\cos^2\theta_{1}\sin^2\theta_{2} + \sin^2\theta_{1}\cos^2\theta_{2}\right) \notag \\
  t_{s} &= \frac{1}{4}\sin 2\theta_{1}\sin 2\theta_{2}\frac{\lambda_{2}\sin\beta}{\lambda_{1}} \label{eq:weakhopping_sin}\\
  t_{c} &= \frac{1}{4}\sin 2\theta_{1}\sin 2\theta_{2}\left(1 - \frac{\lambda_{2}\cos\beta}{\lambda_{1}}\right), \label{eq:weakhopping_cos}
\end{align}
The adjustable parameters in this model are: the ratio of potential strengths \(\lambda_{2}/\lambda_{1}\), the phase difference \(\beta\), the spatial frequency \(\alpha\) and the angles of local unitary transformations \(\theta_{1,2}\).

Model~\eqref{eq:pm} is a special case of a more generic Jacobi operator.
A self-adjoint quasiperiodic Jacobi operator \(J\) acting on \(l^{2}(\mathbb{Z})\) space is the most generic form of a quasiperiodic system,~\cite{teschl2000jacobi}
\begin{align}
    (Ju)_{n} &= v(\alpha n + \phi)u_{n} \\
    & + t_{\rho}(\alpha n + \phi)u_{n+1} + t^{*}_{\rho}(\alpha(n-1) + \phi)u_{n-1}. \notag
\end{align}
Analytical results have been established only for some specific cases of the Jacobi operator.~\cite{aubry1980analyticity, avila2017spectral, jitomirskaya2017arithmetic}
We use these known cases below to characterize the spectrum of the projected model~\eqref{eq:pm} for specific parameters values.

One case admitting an analytical treatment is the Extended Harper model (EHM).
In this case the self-adjoint quasiperiodic Jacobi model is defined as:
\begin{align}
    \label{eq:EHM}
    (Ju)_{n} &= 2\cos(2\pi\alpha n - \pi\alpha)u_{n} \\
    & + 2\rho\left[\cos(2\pi\alpha n)u_{n+1} + \cos(2\pi\alpha (n-1))u_{n-1}\right]. \notag
\end{align}
Both the onsite potential and the hopping are given by the cosine terms with the control parameter \(\rho\).
Structurally this model is similar to our projected model.
The properties of the EHM spectrum have been characterized thoroughly in Ref.~\onlinecite{avila2017spectral}.
For \(2\rho < 1\), all the eigenstates are localized, as guaranteed by the RAGE theorem.~\cite{ruelle1969remark,amrein1973characterization,enss1978asymptotic}
Otherwise, for \(2\rho \geq 1\), the energy spectrum is fractal,~\cite{avila2017spectral} similar to the Aubry--Andr\'e--Harper model at its critical point.~\cite{aubry1980analyticity}
Moreover, the corresponding eigenstates are multifractal, as was demonstrated numerically in Ref.~\onlinecite{chang1997multifractal}.

For infinite \(2\rho\), the onsite potential can be neglected and we are left with the hopping only.
The model in this limit is called the off-diagonal Harper model (OHM),~\cite{han1994critical,kraus2012topological}
\begin{gather}
    \label{eq:OHM}
    (Ju)_{n} = \left[\cos(2\pi\alpha n)u_{n+1} + \cos(2\pi\alpha (n-1))u_{n-1}\right].
\end{gather}
It is self-dual under a modified Fourier transformation similar to that of the Aubry--Andr\'e model.~\cite{lee2023critical}
Consequently the entire spectrum of the model is critical and the eigenstates are fractal.

When the mapping to the extended or the off-diagonal Harper model is not available, we resort to exact diagonalization in order to characterize the properties of the eigenstates.
The inverse participation ratio (IPR)~\cite{wegner1980inverse,castellani1986multifractal,evers2008anderson} for an eigenstate \(\psi_n\) is commonly used to quantify the degree of localization of a wavefunction in a system of size \(L\),
\begin{gather}
    \label{eq:ipr}
    \mathrm{IPR} = \sum_{n}\abs{\psi_{n}}^{4} \sim L^{-\tau},
\end{gather}
IPR scales as a power-law of the system size \(L\) with a scaling exponent \(\tau\).
The exponent \(\tau\) is defined as follows,~\cite{vicsek1992fractal}
\begin{gather}
    \label{eq:tauexplicit}
    \tau = \lim_{L\to\infty}\frac{1}{\ln(1/L)}\ln\operatorname{IPR}.
\end{gather}
However, we cannot use this definition for a single eigenstate as a function of system size \(L\).
Instead, we calculate the average of \(\tau\) over a small energy bin to extract the scaling behavior of \(\tau\).
This is achieved by rescaling the eigenspectrum into the interval \([0, 1]\) and splitting it into equidistant bins \(e\).
This allows us to calculate the average of \(\tau\) in a single bin \(e\), denoted as \(\mbeet\) as a function of lattice size \(L\).
In the thermodynamic limit, all eigenstates in the bin \(e\) are localized if \(\mbeet(L\to\infty) = 0\).
Otherwise, \(\mbeet(L\to\infty)\) takes a finite value between \(0\) and \(1\).

To confirm the localization of an eigenstate, we use the fact \(\tau = 0\) in the thermodynamic limit
and introduce the following linear model for \(\mbeet\), inspired by Eq.~\eqref{eq:tauexplicit}:
\begin{gather}
    \label{eq:linearmodel}
    \mbeet(L) = \frac{\kappa(e)}{\ln(1/L)} + \mathrm{intercept}, \quad \mathrm{intercept} = 0.
\end{gather}
The numerical fit is quantified with the goodness-of-fit \(R^{2}\)~\cite{steel1960principles} which shows how well the data is described by the model~\eqref{eq:linearmodel}.
\(R^2 \approx 1\) indicates that states in bin \(e\) are localized in the thermodynamic limit.
Otherwise, \(R^2 \leq 0\), the set of numerical data does not follow the linear model~\eqref{eq:linearmodel} at all.
Hence, in the thermodynamic limit, there are extended or critical states for which a different fitting model should be used.
This choice of the fitting is motivated by an observation that we only discovered localized or critical states in our previous study~\cite{lee2023critical} and we expect similar results in this extended model.
This expectation has indeed been confirmed a posteriori.

\subsubsection{\(\lambda_{2}/\lambda_{1} = 0\) for any phase difference \(\beta\)}
\label{sec:mappingEHM}

If we only apply the quasiperiodic potential~\eqref{eq:fields} to one leg of the ladder, e.g. for example \(\lambda_{2}/\lambda_{1} = 0\),
then the projected model~\eqref{eq:pm} simplifies: \(\tau_{s}\) is zero and \(v_{s}\) takes the following form,
\begin{gather}
    v_{s} = \sin(\pi\alpha)\cos(\theta_1 - \theta_2)\cos(\theta_2 + \theta_1).
\end{gather}
Then the model~\eqref{eq:pm} maps to the extended Harper model for \(v_{s} \equiv 0\) only.
This produces the following constraint on the angles \(\theta_{1,2}\):
\begin{gather}
    \label{eq:multifractalrelation}
    \theta_{2} + \theta_{1} = \frac{\pi}{2}.
\end{gather}
Using this to eliminate \(\theta_{2}\), we obtain the following expressions for \(v_{c}\) and \(\tau_{c}\)
\begin{align}
    v_{c} &= 2\cos(\pi\alpha)\sin^2\theta_{1}\cos^2\theta_{1},\\
    t_{c} &= \sin\theta_{1}\cos\theta_{1}\sqrt{1 - \cos 4\theta_{1}}/\sqrt{8}.
\end{align}
For these coefficients the model~\eqref{eq:pm} is equivalent to the \(\lambda_2/\lambda_1=0\) case studied in the previous work.~\cite{lee2023critical}. 
Note that for \(\theta_1=0,\pi/2\) both quantities vanish, producing a set of decoupled sites.
Then for \(\theta_1\neq 0,\pi/2\) the control parameter, \(\rho\) of the EHM~\eqref{eq:EHM} is expressed as follows in terms of \(\theta_1\)
\begin{gather}
    \label{eq:commonhopping}
    2\rho = \left|\frac{\sqrt{1 - \cos 4\theta_{1}}/\sqrt{8}}{\cos(\pi\alpha)\sin\theta_{1}\cos\theta_{1}}\right| = \left|\frac{1}{\cos(\pi\alpha)}\right| \geq 1.
\end{gather}
The absolute value above ensures that the negative hopping case is also covered (the sign can be trivially gauged away) and also to match with the definition of the extended Harper model~\cite{avila2017spectral} where non-negative \(2\rho\) is assumed.
Since the model is equivalent to the special case \(\theta_{1,2} = \pi/4\) studied before in Ref.~\onlinecite{lee2023critical} up to a global rescaling,
we have subdiffusive wavepacket spreading for \(\theta_{1,2}\) satisfying Eq.~\eqref{eq:multifractalrelation}.

\begin{figure}
\centering
  \includegraphics[width = \columnwidth]{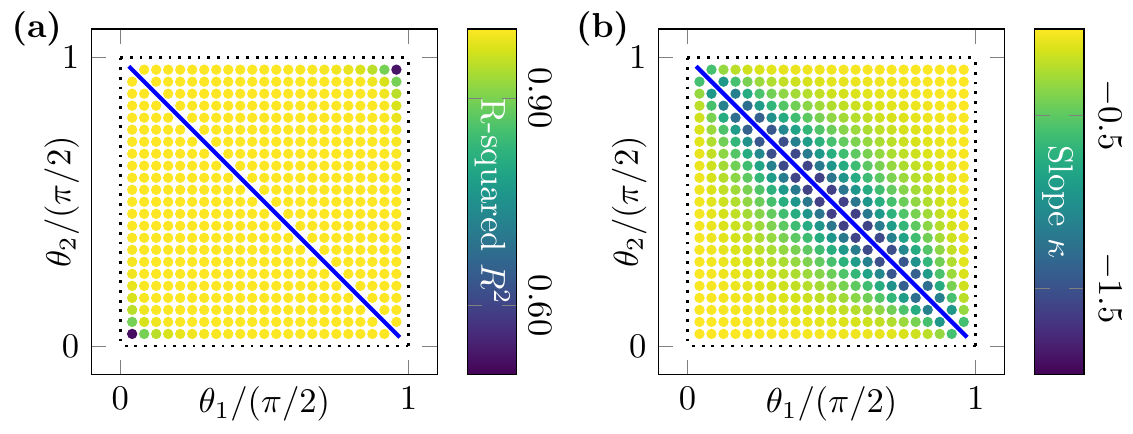}
  \caption{
    Phase diagram \((\theta_1, \theta_2)\) of the projected model~\eqref{eq:pm} for \(\lambda_2/\lambda_1 = 0\), e.g. one of the quasiperiodic fields is absent.
    The yellow dots indicate the fully localized spectrum.
    The diagonal blue line indicates the region of critical spectrum given by Eq.~\eqref{eq:multifractalrelation}.
    The black dotted lines on the border correspond to the case of disconnected dimers: all eigenstates are compactly localized.
    (a) \(R^2\) values for parameters \(\theta_{1},\theta_{2}\): All eigenstates are localized in the thermodynamic limit except the diagonal.
    (b) Slope \(\hat{\kappa}\)~\eqref{eq:linearmodel} for the parameters away from the diagonal of the phase diagram.
  }
  \label{fig:phasediagramEHM}
\end{figure}

For all the other \(\theta_{1,2}\), e.g. not satisfying the above condition~\eqref{eq:multifractalrelation}, we establish the character of the spectrum -- localized or not -- numerically.
We scanned the full parameter region, \(0<\theta_{1,2}<\pi/2\), discretized into the \(25\times 25\) grid.
The exponent \(\mbet\) averaged over the entire spectrum is computed via Eq.~\ref{eq:linearmodel} for lattice sizes \(L = 2000, 4000, \dots 12000\) in steps of \(2000\).
The results are summarized in Fig~\ref{fig:phasediagramEHM}: Apart from the diagonal line, all points have the \(R^2\)-values very close to \(1\).
That is all the eigenstates of the projected model~\eqref{eq:pm} are localized in the thermodynamic limit whenever it does not map onto the extended Harper model, e.g. away from the diagonal \(\theta_1 + \theta_2 = \pi/2\).
The increase of the absolute value of the slope \(\hat{\kappa}\)~\eqref{eq:linearmodel} towards the diagonal line is related to the increase of the localization length as we approach the transition to critical states.
The deviations at the anti-corners of the phase diagram are caused by the failure of the simple linear model~\eqref{eq:linearmodel} to capture the compact localization emerging for \(\theta_{1,2}\to 0\) or \(\theta_{1,2}\to\pi/2\).
We consider these cases separately in Sec.~\ref{sec:mappingCLS} below, and demonstrate that the spectrum is localized in this case: the eigenstates are effectively compactly localized.
We note, that our previous work~\cite{lee2023critical} considered the case of \(\theta_1=\theta_2=\pi/4\).

\subsubsection{\(\lambda_{2} = \lambda_{1}\) and \(\beta = \pi\)}
\label{sec:mappingOHM}

For \(\theta_2=\pi/4\), equal potential strengths \(\lambda_{2} = \lambda_{1} = 1\) and phase difference \(\beta = \pi\), \(v_{s,c} = t_{s} = 0\) and we are left with only the hopping terms \(t_{c} = \sin 2\theta_{1}/2\) in the model~(\ref{eq:pm}, \ref{eq:coefficients:t}).
The projected model becomes exactly equivalent to the off-diagonal Harper model and the role of \(t_{c}\) is simply to rescale the energy:
\begin{gather}
    \label{eq:OHM1}
    Ea_{n} = t_{c}\left[\cos(2\pi\alpha n)a_{n+1} + \cos(2\pi\alpha (n-1))a_{n-1}\right].
\end{gather}
We have established in our previous work, Ref.~\onlinecite{lee2023critical} an almost diffusive spreading of an initially localized wavepacket in the off-diagonal Harper model.
Now the factor \(t_c\) only sets the global timescale for the wavepacket spreading.

\begin{figure}
\centering
  \includegraphics[width = \columnwidth]{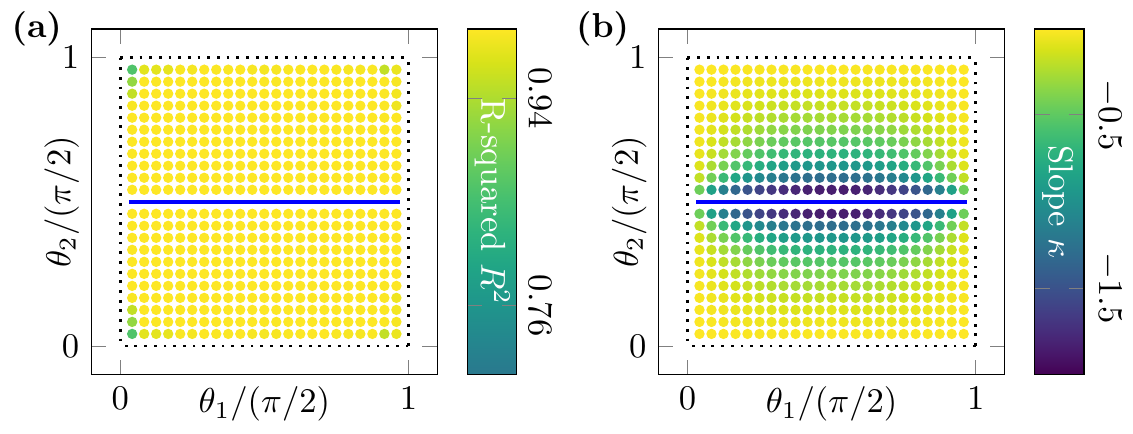}
  \caption{
    Phase diagram \((\theta_1,\theta_2)\) of the projected model~\eqref{eq:pm} for equal strength of the quasiperiodic potentials \(\lambda_2=\lambda_1\) and phase difference \(\beta = \pi\).
    The yellow dots indicate the fully localized spectrum.
    The horizontal blue line, \(\theta_{2} = \pi/4\), the models with critical spectrum, which map to the off-diagonal Harper mode.
    The black dotted lines on the border correspond to the system of disconnected dimers and compactly localized eigenstates.
    (a) \(R^2\) values for parameters \(\theta_{1},\theta_{2}\).
    Aside from the horizontal blue line, all points have the values very close to \(1\): all  eigenstates are localized in the thermodynamic limit.
    (b) Slope \(\hat{\kappa}\) for the parameters away from the horizontal of the phase diagram.
  }
  \label{fig:phasediagramOHM}
\end{figure}

For \(\theta_{2}\neq \pi/4\), we also carried out a numerical scan of the spectrum of the projected model over the region \(0<\theta_{1,2}<\pi/2\), discretized into \(25\times 25\) grid.
The fitting procedure of \(\mbet\) (over the entire spectrum) is performed via Eq.~\ref{eq:linearmodel} for lattice sizes \(L = 2000, 4000, \dots 12000\) in steps of \(2000\).
The observed results follow closely those of the case \(\lambda_2/\lambda_1 = 0\):
all the eigenstates of the projected model~\eqref{eq:pm} are localized in the thermodynamic limit away from the horizontal line of the phase diagram, 
e.g. when the projected model does not map onto the off-diagonal Harper model, as shown in Fig.~\ref{fig:phasediagramOHM}.
The increase of the absolute value of the slope towards the horizontal line is related to the increase of the localization length towards the line of critical states.

\subsubsection{Near zero angles \(\theta_{1,2}\to 0\)}
\label{sec:mappingCLS}

Visible deviations from the at the anti-corners of the phase diagrams are present in Fig.~\ref{fig:phasediagramEHM} and to a lesser extent in Fig.~\ref{fig:phasediagramOHM}.
In this section we focus on the case of \(\theta_{1,2}\to 0\) and demonstrate that all the eigenstates of the projected model~\eqref{eq:pm} are effectively compactly localized.
To see this we approximate \(\sin\theta \approx \theta\) and \(\cos\theta \approx 1\), so that the coefficients~(\ref{eq:coefficients:v}-\ref{eq:coefficients:t}) simplify to:
\begin{align}
    v_{s} &= \sin(\pi\alpha)(\theta_{1}^{2}\theta_{2}^{2} - 1) \\
    &+ \frac{\lambda_{2}}{\lambda_{1}}\cos\beta \sin(\pi\alpha)(\theta_{1}^{2} - \theta_{2}^{2}) - \frac{\lambda_{2}}{\lambda_{1}}\sin\beta \cos(\pi\alpha)(\theta_{2}^{2} + \theta_{1}^{2}) \notag \\
    v_{c} &= \cos(\pi\alpha)(1 + \theta_{1}^{2}\theta_{2}^{2}) \\
	&+ \frac{\lambda_{2}}{\lambda_{1}}\sin\beta\sin(\pi\alpha)(\theta_{1}^{2} - \theta_{2}^{2}) + \frac{\lambda_{2}}{\lambda_{1}}\cos\beta\cos(\pi\alpha)(\theta_{2}^{2} + \theta_{1}^{2}) \notag \\
    t_{s} &= \theta_{1}\theta_{2}\frac{\lambda_{2}\sin\beta}{\lambda_{1}} \\
    t_{c} &= \theta_{1}\theta_{2}\left(1 - \frac{\lambda_{2}\cos\beta}{\lambda_{1}}\right)
\end{align}
Keeping only linear terms in \(\theta_{1,2}\), so that \(\theta_{1}^{2} = \theta_{2}^{2} = \theta_{1}\theta_{2} = 0\), 
we see that the hoppings all vanish \(t_{s} = t_{c} = 0\) and only the onsite potential terms are left.
This implies that all the eigenstates are effectively compactly localized but their energies are non-degenerate, because of the onsite potential.
A similar argument implies that all states are compactly localized for \(\theta_{1,2} \approx \pi/2\).

\section{Finite perturbation}
\label{sec:finite}

We considered so far the case of weak quasiperiodic potential and found critical eigenstates for specific angles of the local unitary transformations, specific potential strengths \(\lambda_{1,2}\) and the phase difference \(\beta\).
Naturally, the next question is what happens at finite perturbation strength.
Generically, we expect that eigenstates localized for the weak potential remain localized upon increasing the potential strength, and that strong enough potential, as compared to the bandgap \(\Delta\), should localize all the eigenstates.
The open question is what happens to the critical eigenstates for moderate values of \(\lambda_{1,2}\)?

To address this issue, it is convenient to work with the \emph{semi-detangled} Hamiltonian,~\cite{danieli2021nonlinear,lee2023critical}
which is defined by inverting only the second unitary transformation \(U_{2}\) (see Sec.~\ref{sec:model}).
This defines the semi-detangled basis \(\{u_{n}, d_{n}\}\) and gives the new, semi-detangled Hamiltonian,
\begin{gather}
    \label{eq:ham-sd}
    \mhsd = U_{1} \mhfd U_{1}^{\dagger} + U_{2}^{\dagger}WU_{2}.
\end{gather}
The semi-detangled wavefunction amplitudes \(u_{n}\) and \(d_{n}\) are related to the basis states of the ABF Hamiltonian \(\{p_{n}, f_{n}\}\) as follows,
\begin{align}
    u_{n} &= p_{n}\cos\theta_{2} - f_{n}\sin\theta_{2} \notag \\
    d_{n} &= p_{n}\sin\theta_{2} + f_{n}\cos\theta_{2}.
\end{align}
The model~\eqref{eq:ham-sd} takes the following form in the semi-detangled basis:
\begin{align}
    \label{eq:semi}
    Eu_{n} &= \left[\varepsilon_{b}\sin^{2}\theta_{1} + \varepsilon_{a}\cos^{2}\theta_{1}\right] u_{n} \notag \\
    &+ \left[W_{2}(n)\sin^{2}\theta_{2} + W_{1}(n)\cos^{2}\theta_{2}\right] u_{n}  \\
    &+ \left[(W_{1}(n) - W_{2}(n))\cos\theta_{2}\sin\theta_{2}\right] d_{n} +\left[\Delta\cos\theta_{1}\sin\theta_{1}\right] d_{n+1}, \notag \\
    Ed_{n} &= \left[\varepsilon_{a}\sin^{2}\theta_{1} + \varepsilon_{b}\cos^{2}\theta_{1}\right] d_{n} \notag \\
    &+ \left[W_{1}(n)\sin^{2}\theta_{2} + W_{2}(n)\cos^{2}\theta_{2}\right] d_{n} \\
    &+ \left[(W_{1}(n) - W_{2}(n))\cos\theta_{2}\sin\theta_{2}\right] u_{n} + \left[\Delta\cos\theta_{1}\sin\theta_{1}\right] u_{n-1}. \notag
\end{align}
This model is at the core of the subsequent analytical and numerical analysis.

\subsubsection{Perturbation independent fractality edges for \(\lambda_{2} = 0\)}
\label{sec:claimEHM}

We remove one of the quasiperiodic potentials, \(\lambda_2 = 0\) and set \(\theta_2 = \pi/2 - \theta, \theta_1 = \theta\).
The model in the semi-detangled basis takes the following form,
\begin{align}
    \frac{E u_{n}}{\cos\theta\sin\theta} &= \left[\frac{\varepsilon_{a}}{\tan\theta} + \frac{\left(\varepsilon_{b} + W_{1}(n)\right)}{1/\tan\theta}\right]u_{n} \notag \\
    &+ W_{1}(n)d_{n} + \Delta d_{n+1}, \label{eq:semiFE1} \\
    \frac{E d_{n}}{\cos\theta\sin\theta}  &= \left[\frac{\left(\varepsilon_{b} + W_{1}(n)\right)}{\tan\theta} + \frac{\varepsilon_{a}}{1/\tan\theta}\right]d_{n} \notag \\
    &+ W_{1}(n)u_{n} + \Delta u_{n-1}. \label{eq:semiFE2}
\end{align}
After multiplying Eq.~\eqref{eq:semiFE2} by \(\tan\theta\) and taking the difference of the above equations,
we get a single equation without quasiperiodic potentials:
\begin{gather}
    E_{u}(E,\theta) u_{n} + \tan\theta E_{g}u_{n-1} = \tan\theta E_{d}(E,\theta)d_{n} + E_{g}d_{n+1}, \notag
\end{gather}
where \(E_{u}(E,\theta)\) and \(E_{d}(E,\theta)\) are
\begin{align}
    E_{u}(E,\theta) &= \left[\frac{E}{\cos\theta\sin\theta} - \left(\frac{E_{a}}{\tan\theta} + E_{b}\tan\theta\right)\right] \\
    E_{d}(E,\theta) &= \left[\frac{E}{\cos\theta\sin\theta} - \left(\frac{E_{b}}{\tan\theta} + E_{a}\tan\theta\right)\right]
\end{align}
For \(\theta_{1,2} = \pi/4\), we obtain the model discussed in Ref.~\onlinecite{lee2023critical} (\(\sigma = \varepsilon_{a} + \varepsilon_{b}\)).
There are fractality edges at the flatband energies \(\varepsilon_{a,b}\) for any perturbation strength \(\lambda_1\) and all the eigenstates in between are critical:
\begin{gather}
    (2E - \sigma)u_{n} + E_{g}u_{n-1} = (2E - \sigma)d_{n} + E_{g}d_{n+1}.
\end{gather}
This result can be rationalized with a simple assumption that the critical states appear when the hopping is larger than the onsite potential:
\begin{gather}
  \label{eq:ineq}
  |2E - \sigma| \leq E_{g}.
\end{gather}
implying \(\varepsilon_{a} \leq E \leq \varepsilon_{b}\) for the critical states.

\begin{figure}
\centering
  \includegraphics[width = \columnwidth]{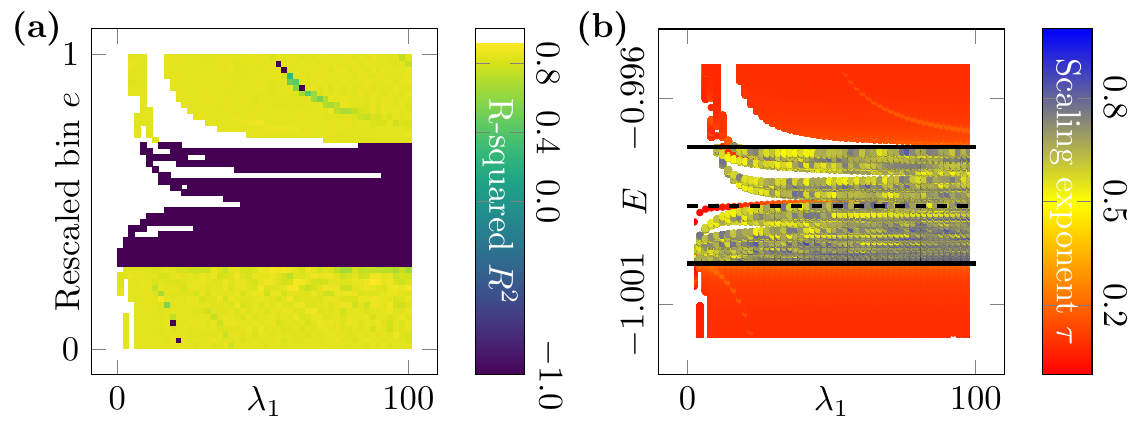}
  \caption{
    Fractality edges at finite potential strength \(\lambda_1\) and \(\lambda_{2} = 0\), for \(\theta = 0.01\pi/2\) and lattice size \(L = 10946\).
    (a) Fractality edges in the rescaled spectrum.
    The negative \(R^{2}\) values are replaced with \(-1\) for better visual separation of localized and critical regions.
    (b) Fractality edges in the original, non-rescaled energy spectrum.
    The bottom black line is the flatband \(\varepsilon_{a} = -1\), while the top black black is \(E \approx -0.997\).
    The dashed black line is the upper bound \(E \approx -0.998\) given by Eq.~\eqref{eq:inequalities}.
  }
  \label{fig:EHM0P01}
\end{figure}

\begin{figure}
\centering
  \includegraphics[width = \columnwidth]{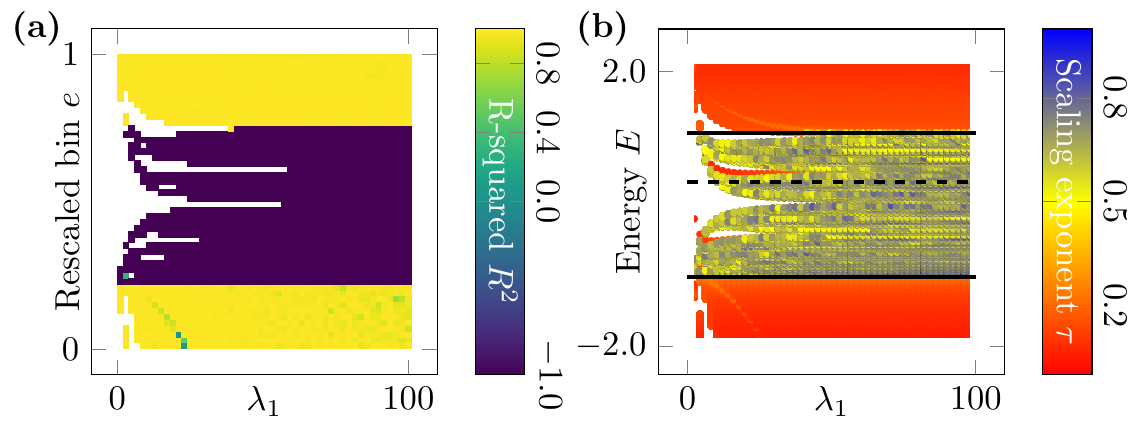}
  \caption{
    Fractality edges at finite potential strength \(\lambda_1\) and \(\lambda_{2} = 0\), for \(\theta = 0.68\pi/2\) and lattice size \(L = 10946\).
    (a) Fractality edges in the rescaled spectrum.
    The negative \(R^{2}\) values are replaced with \(-1\) for better visual separation of localized and critical regions.
    (b) Fractality edges in the original, non-rescaled energy spectrum.
    The bottom black line is the flatband \(\varepsilon_{a} = -1\), while the top black line is \(E \approx 1.05\).
    The dashed black line is the upper bound \(E \approx 0.393\) given by Eq.~\eqref{eq:inequalities}.
  }
  \label{fig:EHM0P68}
\end{figure}

For \(\theta \neq \pi/4\), following the same logics, the above inequailty for energy \(E\) is modified as follows:
\begin{gather}
    \label{eq:inequalities}
    |E_{u}| < E_{g}\tan\theta \quad\text{and}\quad |E_{d}|\tan\theta < E_{g}.
\end{gather}
However, now the simple constraint does not explain the entire range of fractality edges but only covers the narrower range of the critical states as seen in Fig.~\ref{fig:EHM0P01}-\ref{fig:EHM0P68}.

Examples are shown and summarized in Fig.~\ref{fig:EHM0P01} and~\ref{fig:EHM0P68}, where \(\theta = 0.01\pi/2\) and \(\theta = 0.68\pi/2\), respectively.
We fixed the flatband energies to \(\varepsilon_{a} = -1\) and \(\varepsilon_{b} = 2\).
Since the fractality edges are potential strength independent, we only considered part of the spectrum around the edges, and used it for the rescaled spectrum.
We apply the linear model introduced in Eq.~\eqref{eq:linearmodel}, and introduce \(50\) energy bins \(e\) for the spectrum rescaled to fit in \([0, 1]\).
The lattice sizes (in unit cells) are \(L = 3283, 4181, 4832, 5473, 6765, 10946\).
The exact model has two sites for each unit cell and the size of the Hamiltonian matrix is \(2L\times 2L\).
In both examples, there are two fractality edges independent of the potential strength \(\lambda_1\), with critical states in between and localized states outside.
The lower fractality edge coincides with the flatband energy \(\varepsilon_{a}\).
By symmetry, setting \(\lambda_1=0\) and varying \(\lambda_2\), one would get the similar spectral behavior with the upper fractality edge would be at \(\varepsilon_{b}\).

\subsubsection{Existence of CIT at \(\lambda_{2} = \lambda_{1}\) and \(\beta = \pi\)}
\label{sec:claimOHM}

Let \(\theta_{1} = \theta\), \(\theta_{2}=\pi/4\), \(\lambda_2=\lambda_1=\lambda\) and the phase difference \(\beta = \pi\), hence \(W_{2} = -W_{1} = -W\).
Then the model~\eqref{eq:semi} simplifies as follows:
\begin{align}
    E_{u}(E,\theta)u_{n} &= W(n)d_{n} + \Delta\cos\theta\sin\theta d_{n+1}, \\
    E_{d}(E,\theta)d_{n} &= W(n)u_{n} + \Delta\cos\theta\sin\theta u_{n-1},
\end{align}
where \(E_{u}(E,\theta)\) and \(E_{d}(E,\theta)\) are defined as,
\begin{align}
    E_{u}(\theta) &= E - \varepsilon_{b}\sin^{2}\theta - \varepsilon_{a}\cos^{2}\theta, \notag \\
    E_{d}(\theta) &= E - \varepsilon_{a}\sin^{2}\theta - \varepsilon_{b}\cos^{2}\theta. \notag
\end{align}
We can eliminate \(u_{n}\) from the above equations by substitution, resulting in equations with \(d_n\) only
\begin{align}
    E_{d}(\theta)E_{u}(\theta)d_{n} &= W^{2}(n)d_{n} + \Delta^{2}\cos^{2}\theta\sin^{2}\theta d_{n}, \\
    &+ \Delta\cos\theta\sin\theta\left[W(n)d_{n+1} + W(n-1) d_{n-1}\right]. \notag
\end{align}
This can be further recast into the model considered in our previous work~\cite{lee2023critical}:
\begin{align}
    \label{eq:semiCIT}
    \tilde{E}(\theta, \lambda) d_{n} &= \cos(4\pi\alpha n)d_{n} \\
    &+ K(\theta, \lambda)) \left[\cos(2\pi\alpha (n-1))d_{n-1} + \cos(2\pi\alpha n)d_{n+1}\right], \notag
\end{align}
where \(\tilde{E}(\theta,\lambda)\) and \(K(\theta,\lambda)\) are defined as,
\begin{align}
    \tilde{E}(\theta, \lambda) &= \frac{2}{\lambda^{2}}E_{d}(\theta)E_{u}(\theta) - \frac{1}{2}, \notag \\
    K(\theta,\lambda) &= \frac{2\Delta}{\lambda}\cos\theta\sin\theta. \notag
\end{align}
As discussed in Ref.~\onlinecite{lee2023critical}, the transition between localized and critical states occurs for \(K(\theta,\lambda) = 1\):
\begin{gather}
    \label{eq:CIT}
    \lambda_{c} = \Delta\sin 2\theta.
\end{gather}

\begin{figure}
\centering
  \includegraphics[width = \columnwidth]{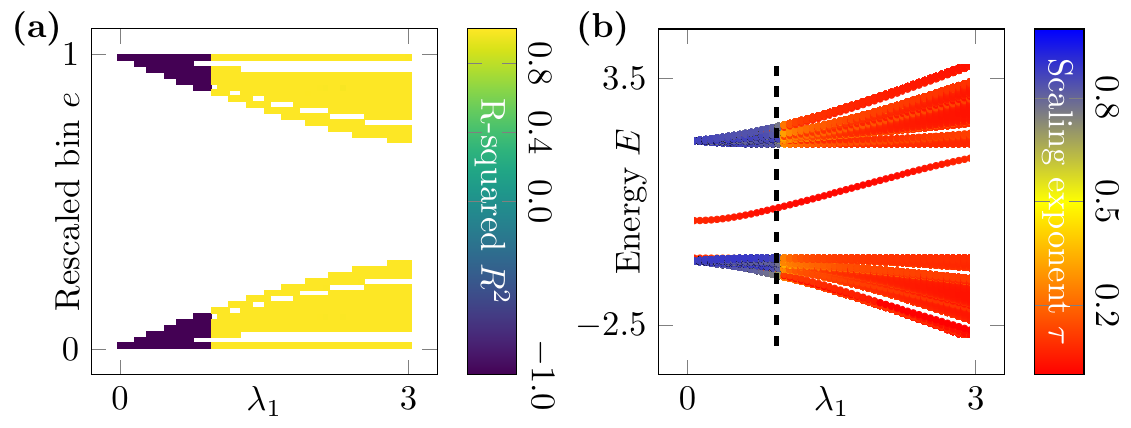}
  \caption{
    Critical-to-insulator transition at finite potential strength \(\lambda_{2} = \lambda_{1}\) for lattice size \(L = 10946\), \(\beta = \pi\) at \(\theta = 0.1\pi/2\).
    (a) Rescaled spectrum.
    The negative \(R^{2}\) values are replaced with \(-1\) for better visual separation of localized and critical regions.
    (b) Original, non-rescaled energy spectrum.
    The vertical dashed black line marks the CIT transition predicted by Eq.~\eqref{eq:CIT}.
  }
  \label{fig:OHM0P1}
\end{figure}

An example for \(\theta = 0.1\pi/2\) is shown and summarized in Fig.~\ref{fig:OHM0P1}.
The flatband energies are taken to be \(\varepsilon_{a} = -1\) and \(\varepsilon_{b} = 2\).
Entire spectrum is rescaled for each \(\lambda_{1}\). 
Then we apply the linear model introduced in Eq.~\eqref{eq:linearmodel} introducing \(50\) energy bins \(\tilde{e}\) for the spectrum rescaled to fit in \([0, 1]\).
Lattice sizes (in unit cells) are \(L = 3283, 4181, 4832, 5473, 6765, 10946\).
The exact model has two sites for each unit cell, the size of the Hamiltonian matrix is \(2L\times 2L\).
The transition point \(\lambda_{c} \approx 0.927\) based on Eq.~\eqref{eq:CIT} is indicated with the black dashed line.
It matches perfectly with the numerical results.
All states are critical to the left of \(\lambda_{c}\) and localized to the right of it.
The string of eigenstates in between the two broadened flatband energies is due to the open boundary conditions.

\section{Conclusions}
\label{sec:conclusion}

We studied the effect of quasiperiodic perturbation, composed of two potentials, on a general model of the ABF manifold with two bands, extending our previous work, 
Ref.~\onlinecite{lee2023critical}, that focused on the specific point \(\theta_{1,2} = \pi/4\) of the ABF manifold.
First we considered the case of weak quasiperiodic perturbation.
We identified the ABF submanifolds of models with spectra containing critical states, and exhibiting subdiffusive and almost diffusive transport for weak quasiperiodic perturbation.
These submanifolds are the direct extension of the cases found in our previous work and can be mapped onto the same extended Harper model.
Namely, the critical states with subdiffusive transport were found for the ABF models satisfying the condition \(\theta_{1} + \theta_{2} = \pi/2\) for the angles of the unitary transformation generating the model.
The other constraint is the absence of one of the two quasiperiodic fields.
The other submanifold with models featuring critical states with almost diffusive transport is given by \(\theta_{2} = \pi/4\) and arbitrary \(\theta_1\).
The quasiperiodic potentials must have equal strengths and the relative phase \(\beta = \pi\), e.g. be of opposite signs.
All such models can be mapped onto the off-diagonal Harper model.
Outside of these manifolds, all the states are localized in the thermodynamic limit as suggested by numerical analysis.
Specifically, for small angles \(\theta_{1,2}\) we demonstrated compact localization of the spectrum.

For a finite quasiperiodic potential, the perturbed ABF models which map to the extended Harper model, display fractality edges in the spectrum separating critical from localized states.
These edges are independent of the potential strength.
On the other hand, the models which map onto the off-diagonal Harper model, show no fractality edges.
Instead they exhibit a critical-to-insulator transition.
The transition point is derived analytically and confirmed numerically, and depends on the angle \(\theta_1=\theta_2=\theta\).

Only localized and critical eigenstates appeared for the type of onsite quasiperiodic potential that we have considered.
An interesting problem is to identify other quasiperiodic potentials that might give delocalized states in projected models, for example by mapping the ABF model for weak perturbation onto the extended Harper model in the delocalized part of the phase diagram.
This could be achieved by solving an inverse problem of reconstructing the full potential from the projected effective model.

\begin{acknowledgments}
  This work was supported by the Institute for Basic Science, Project Code (Project No.~IBS-R024-D1).
\end{acknowledgments}


\bibliography{general,flatband,frustration,mbl}

\end{document}